\documentclass[
 aip,
 amsmath,amssymb,
 reprint,%
]{revtex4-1}

\usepackage{graphicx}% Include figure files
\usepackage{dcolumn}% Align table columns on decimal point
\usepackage{bm}% bold math
\usepackage[utf8]{inputenc}
\usepackage[T1]{fontenc}
\usepackage{mathptmx}
\usepackage{etoolbox}
\usepackage{amsmath}
\usepackage{mathtools}

\makeatletter
\def\@email#1#2{%
 \endgroup
 \patchcmd{\titleblock@produce}
  {\frontmatter@RRAPformat}
  {\frontmatter@RRAPformat{\produce@RRAP{*#1\href{mailto:#2}{#2}}}\frontmatter@RRAPformat}
  {}{}
}%
\makeatother

\newcommand{\tder}[1]{\frac{d #1}{dt}} % time derivative
\DeclarePairedDelimiter{\norm}{\lVert}{\rVert}

\begin{document}

\preprint{AIP/123-QED}

\title{Sparse identification of effective microparticle interaction potential in dusty plasma from simulation data}
% Force line breaks with \\
\author{Z.~B.~Howe}
\email{zbrookshowe@auburn.edu}
\affiliation{Department of Physics, Auburn University, Auburn, AL, USA}

\author{L.~S.~Matthews}

\author{T.~Hyde}
\affiliation{Center for Astrophysics, Space Physics, and Engineering Research (CASPER), Baylor University, Waco, TX, USA}

\author{L.~Guazzotto}

\author{E.~G.~Kostadinova}
\affiliation{Department of Physics, Auburn University, Auburn, AL, USA}

% \author{C. Author}
%  \homepage{http://www.Second.institution.edu/~Charlie.Author.}
% \affiliation{%
% Second institution and/or address%\\This line break forced% with \\
% }%

\date{\today}% It is always \today, today,

\begin{abstract}
Identification of the particle interaction potential is a challenging and important task in dusty plasma, colloids, and smart materials as it allows the characterization of structure formation and helps predict phase transitions. 
With the advent of machine learning methods, this interaction can be extracted from particle position data, leading to a generalizable expression which is applicable in different systems. 
Methods such as sparse regression aim to provide a physically interpretable model that can generalize well, while avoiding unnecessary complexity due to overfitting. 
In this work, we present the use of the Sparse Identification of Nonlinear Dynamics (SINDy) with the weak formulation to learn equations of motion for noisy data from simple simulations of two dust particles interacting with a Yukawa (shielded Coulomb) potential.
The application of these methods to experimental dusty plasma data is discussed, particularly in the case of simulation data and glass box experiments in RF discharge gravity environments and DC discharge microgravity environments, such as the Plasmakristall-4 (PK-4) experiment.
\end{abstract}

\maketitle

% \begin{quotation}
% The ``lead paragraph'' is encapsulated with the \LaTeX\ 
% \verb+quotation+ environment and is formatted as a single paragraph before the first section heading. 
% (The \verb+quotation+ environment reverts to its usual meaning after the first sectioning command.) 
% Note that numbered references are allowed in the lead paragraph.
% %
% The lead paragraph will only be found in an article being prepared for the journal \textit{Chaos}.
% \end{quotation}

%%%%%%%%%%%%%%%%%%%%%%%%%%%%%%%%%%%%%%%%%%%%%%%%%%%%%
% INTRODUCTION
%%%%%%%%%%%%%%%%%%%%%%%%%%%%%%%%%%%%%%%%%%%%%%%%%%%%%
\section{\label{sec:intro}Introduction}
Plasmas, and particularly dusty plasmas, are abundant on earth and in space. 
Dusty plasmas are a four-component system consisting of ions, electrons, neutral gas particles, and macroscopic ``dust particles,'' typically ranging from $10^{-9}$ m to $10^{-6}$ m in diameter for laboratory experiments. 
Fundamentally, dusty plasmas represent a macroscopic model of strongly interacting complex systems as they have been shown to exhibit interesting collective effects, including structure formation, waves and instabilities, and turbulence  \cite{beckers_physics_2023}. 

Dust particle flows provide a unique window into the study of nonlinear waves \cite{merlino_nonlinear_2012,shukla_nonlinear_2003}, while the dynamic dust motion can give insight into the thermodynamics of the system and the properties of the background plasma discharge \cite{andrew_anisotropic_2025-1,land_glow_2013,scott_mapping_2019}.
Collections of dust particles can also exhibit lattice structures, which can undergo phase transitions \cite{klumov_structural_2009,jaiswal_melting_2019,hartmann_crystallization_2010,singh_experimental_2023}. 
Since the constituent particles are macroscopic and visible at the single particle (kinetic) level, dusty plasmas allow unique insight into these fluid dynamics and condensed matter phenomena. 
Generally, dusty plasma monolayer structures are studied with ground-based experiments whereas microgravity environments allow for the the generalization of such studies to three-dimensional structures.

Of particular interest are the three-dimensional structures that can be observed in microgravity laboratory environments on board the International Space Station (ISS). The first of these, PKE-Nefedov, allowed the first observation of body-center cubic structures in the dust cloud \cite{nefedov_pke-nefedov_2003}. 
The successor of this laboratory, PlasmaKristall-3 (PK-3) Plus \cite{thomas_complex_2008}, offered the first insight into the so-called electrorheological (ER) plasma state \cite{ivlev_first_2008,ivlev_electrorheological_2010}, which is characterized by the ability to tune interparticle potentials by applying an external electric field, which induces a dipole-like interaction between neighboring dust grains.
This implies the existence of an anisotropy in the dust-dust interaction potential causing deviations from the spherically symmetric Yukawa (or shielded Coulomb) potential, which normally leads to an isotropic dust structure without the presence of the external field.
More recently, the Plasmakristall-4 (PK-4) experiment \cite{pustylnik_plasmakristall-4_2016} (currently on board the ISS) has shown similar string-like dust structures reminiscent of those observed on PK-3 Plus, also implying the existence of an anisotropic potential. 
Owing to the different geometry and discharge type of PK-4 compared to PK-3, the dust-dust interaction potential and corresponding structure observed in the PK-4 experiment may differ from the aforementioned ER plasma state \cite{mitic_long-term_2021, vermillion_interacting_2024}.
Namely, the observed string-like structures in PK-4 exhibit anisotropic pair correlations and diffusion properties, suggesting crystalline-like coupling within filaments and liquid-like coupling across filaments \cite{kostadinova_liquid_2023, gehr_structural_2025,andrew_anisotropic_2025-1,andrew_anisotropic_2025-2}. 
In addition, these structures have been recently shown to exhibit anisotropic anomalous diffusion \cite{andrew_anisotropic_2025-1,andrew_anisotropic_2025-2}. 

The physical phenomenon leading to the existence of these anisotropies in the dust-dust interaction is the presence of ion wakes that arise due to external electric fields in the plasma. 
The ion wakes can be thought of as regions of positive space charge forming downstream of the dust particle in the direction of the electric-field-induced ion flow. 
These regions of positive charge attract neighboring negatively charged dust grains, competing with the repulsive dust-dust interaction.
It has been proposed that the dust alignment in filamentary structures in PK-4 is further enabled by the presence of high-frequency discharge ionization waves that affect the ion streaming velocity and corresponding wakefield structure \cite{hartmann_ionization_2020, matthews_effect_2021}.

In addition to experimental observations, the effect of ion wakefields on dust-dust interactions has been modeled both analytically and computationally.
An investigation using kinetic theory, considering both collisions and ion drift, led to the derivation of an analytic form for the potential around a dust grain \cite{kompaneets_potential_2007, kompaneets_wakes_2016}
\begin{equation}\label{eq:dipole_potential} 
    \varphi = \frac{q_d}{4\pi\varepsilon_0r}\left[
        e^{-r/\lambda_{Di}}
        - 0.43\frac{M^2\lambda_{Di}^2}{r^2}(3\cos^2\zeta-1)
    \right],
\end{equation}
where $q_d$ is the dust charge, $\lambda_{Di}$ is the ion Debye length, $M$ is the thermal Mach number of the ion drift velocity, and $\zeta$ is the angle between the external electric field and the vector connecting two particles. The first term is the isotropic Yukawa potential and the second term is an effective dipole interaction. This expression has been applied in analysis of various experimental environments, including the PK-3 Plus experiment \cite{ivlev_electrorheological_2010} as well as the PK-4 experiment \cite{pustylnik_plasmakristall-4_2016,vermillion_interacting_2024,mitic_long-term_2021,pustylnik_three-dimensional_2020}.
While the form of the expression was found to predict the ER dust structure observed in PK-3 well, the model was not shown to be as predictive of structure formation in the dust cloud of PK-4 \cite{mendoza_ion_2025,vermillion_modeling_2022}.
The present project aims to build on previous analytical and numerical work by learning the functional form for the dust-dust interaction potential directly from data using a physics-informed machine learning approach.

The remainder of this paper is organized as follows. 
In Sec.~\ref{sec:theory} we outline the relevant physics behind dust interactions in dusty plasma and outline the use of machine learning or data-driven methods in plasma physics. 
In Sec.~\ref{sec:met}, We present the Sparse Identification of Nonlinear Dynamics (SINDy) method in the weak formulation, explain the manner in which synthetic data are generated, and discuss the cross-validation used here.
In Sec.~\ref{sec:res}, we show the results using weak SINDy with cross-validation to identify equations of motion from synthetic data.
In Sec.~\ref{sec:dis}, the results are interpreted and limitations and future applications are discussed, and finally in Sec.~\ref{sec:con} the conclusions are presented.

\section{\label{sec:theory} Theoretical Background}

\subsection{\label{ssec:ion-wake} Ion-wake-mediated interactions in dusty plasma}

The ion dynamics in PK-4 have also been investigated computationally through simulations of the relevant plasma discharge with and without dust. 
In Ref.~\onlinecite{hartmann_ionization_2020}, the presence of ionization waves in the PK-4 DC discharge was predicted from a Particle In Cell with Monte Carlo Collisions (PIC/MCC) simulation and discovered experimentally using a ground-based version of PK-4 at Baylor University. 
Ionization waves are associated with large spikes in the electric field that occur at \textasciitilde 10 kHz, and further investigation has shown that this strongly influences the tendency of the dust cloud to form string-like structures \cite{vermillion_influence_2022,mendoza_ion_2025}. 
Furthermore, it has been shown that the position of dust particles with respect to neighboring particles affects the grain charge and shape of the ion wake \cite{matthews_dust_2020, vermillion_interacting_2024}.
This breaks the assumption of an isolated dust particle in a sheath made in the kinetic model described by Eq.~\ref{eq:dipole_potential} \cite{kompaneets_potential_2007}.
Approaching this question with a bottom-up (data-driven) approach could provide new insights into the problem that can supplement the top-down (analytical, computational) approach. 

An important aspect of ion-wake-mediated interactions between neighboring dust particles is that when the background ion flow is unidirectional, ion wakes appear only on one side of the particles. This leads to nonreciprocal dust-dust interactions, i.e. $F_{12}\ne F_{21}$, where $F_{ij}$ is the force on particle $i$ by particle $j$. 
This effect has been studied experimentally \cite{yu_physics-tailored_2025, lisin_experimental_2020, melzer_dust_2019}, computationally \cite{matthews_dust_2020}, and theoretically \cite{ivlev_statistical_2015}. 
In the PK-4 experiments, this effect is lessened by symmetric polarity switching of the external electric field, which leads to the formation of ion wakes on both sides of the dust particle along the direction of the electric field.  
In these conditions, the interaction along the direction of the electric field is expected to be reciprocal. However, the 3D interaction potential is expected to be anisotropic---with measurable difference between the field-aligned versus cross-field interaction components.

\subsection{\label{ssec:ML} Data-driven discovery of equations}
Recent developments in the analysis of scientific data have jump-started a revolution in the investigation of natural processes, introducing the idea of the ``data-driven discovery'' of natural laws using optimization algorithms. 
This ``learning'' of governing equations directly from data can fall under the umbrellas of statistical or machine learning, whereas artificial intelligence can be defined as the \textit{usage} of these methods to achieve the automation of scientific discovery. 
Data-driven discovery has been termed by some to be the ``fourth paradigm of scientific investigation,'' in addition to empirical experimentation, analytical derivation, and computational investigation \cite{brunton_data-driven_2022,hey_fourth_2009}. 
The key idea here is to use optimization to extract underlying patterns within scientific data to discover or construct a new description of the physical system of interest. 
This is contrasted against the aforementioned ``first three'' paradigms of scientific research, which seek to observe the natural system (experimentation) and use a first-principles approach to predict or arrive at an understanding of the system (derivation and computation). 
Most of the works referenced in this article utilize one or multiple of these three approaches, but given the nature of data from experiments (video data of dust particles), dusty plasma lends itself well to the approach of deducing an equation describing the dynamics of the dust particles directly from data using optimization.

Within the field of dusty plasma, data-driven approaches have begun to be applied in various contexts. Typically, this involves an artificial neural network (ANN) to measure some properties of the dust dynamics or background plasma. 
Refs.~\onlinecite{liang_determining_2023,liang_full_2024} use an ANN trained on Langevin simulation data of a two-dimensional dusty plasma to recover the dust particle screening length (Debye length) and Coulomb coupling parameter. Refs.~\onlinecite{yu_extracting_2022,yu_physics-tailored_2025} use a neural network approach to learn non-reciprocal dust-dust interaction forces in a three-dimensional experimental dusty plasma system with up to 18 particles. 
From this, the authors are able to back out plasma parameters such as the relevant screening (Debye) length of pairs of particles as well as particles' mass and charge \textit{in situ}. 
ANNs have also been used to perform particle tracking \cite{wang_microparticle_2020,melzer_three-dimensional_2025} and to characterize a phase transition between the string-like and isotropic dusty plasma states \cite{dietz_phase_2021}. 
There have also been data-driven analyses conducted without the use of ANNs. 
For instance, Ref.~\onlinecite{huang_identification_2019} uses a Support Vector Machine to identify an interface between different sizes of dust particles in a binary dusty plasma. 
In an approach similar to that which will be presented here, Ref.~\onlinecite{ding_machine_2021} uses a Bayesian regression to learn the equation of motion of a single dust particle in a glass box. 
In this paper, however, the form of the equation was determined before the regression was performed, which is different than the sparse regression used here, where the algorithm itself determines the form of the equation.

An alternative to the commonly used ANN approach is to use a linear regression to learn equations from data. 
This class of approaches has the distinct advantage of being much less computationally intensive, as well as being more interpretable. 
While the ANN is often referred to as a ``black box,'' regression methods are transparent in the way the solutions are determined. 

Recently, the idea of sparsity-promotion in optimization schemes has led to the more widespread use of ``sparse regression'', which has found a place in many plasma physics applications. 
Sparse regression employs a linear least-squares regression with the presence of a sparsity prior, which forces the optimization scheme to balance minimizing the error with minimizing the complexity of the learned statistical model. 
This is related to Occam's Razor, which can be stated as ``the simplest explanation is usually the best one.'' 
These ideas and others relating to data-driven approaches are discussed extensively in Ref.~\onlinecite{brunton_data-driven_2022}. 

Within plasma physics, sparsity-promoting algorithms have been used in three broad senses: system identification, sparse sensing, and gaussian process regression. 
A complete discussion of the ways that sparse regression have been used for plasma can be found in Ref.~\onlinecite{kaptanoglu_sparse_2023} with an extensive list of works using these methods. 
In the present article, we focus on the Sparse Identification of Nonlinear Dynamics (SINDy) method \cite{brunton_discovering_2016}, described more fully in Sec.~\ref{sec:met}. 
In plasma physics, SINDy has been used notably in stellarator optimization \cite{kaptanoglu_optimization_2025}, to identify a predator-prey model for flows in a tokamak \cite{dam_sparse_2017}, to characterize anode glow oscillations in a low temperature discharge \cite{thakur_data_2022}, and to discover fluid closures and other reduced-order models from plasma simulations \cite{alves_data-driven_2022}, to name a few. 
To the best of our knowledge, the SINDy method has not been applied within the field of dusty plasma.

%%%%%%%%%%%%%%%%%%%%%%%%%%%%%%%%%%%%%%%%%%%%%%%%%%%%%
% METHODS
%%%%%%%%%%%%%%%%%%%%%%%%%%%%%%%%%%%%%%%%%%%%%%%%%%%%%
\section{\label{sec:met}Methods}
In this article, we seek to validate the use of the Sparse Identification of Nonlinear Dynamics (SINDy) method to identify an equation of motion from trajectories of interacting particles, which is of particular interest to those who study dusty plasma.
Since the dust grains in many experiments are of sufficient macroscopic size, their trajectories can be recovered using methods such as particle tracking or particle image velocimetry \cite{thomas_benchmarking_2010,williams_application_2011}.
We show here that the use of particles' individual position and velocity measurements as a time series can be used in a linear regression problem that can uncover individual particle equations of motion given a certain basis of nonlinear (not necessarily linearly indepedent) functions.
The SINDy method picks only the most important terms from this basis, leading to a parsimonious and interpretable equation of motion that can provide insight into the physical mechanisms underlying the dynamics in a dusty plasma system.

This section is organized as follows.
In Sec.~\ref{ssec:sindy}, the SINDy method is presented in its original formulation, along with the sparsity-promoting algorithm, the sequentially thresholded least squares (STLSQ), and the weak formulation of the problem. Then, a simple isotropic Yukawa simulation is presented in Sec.~\ref{ssec:iso}, with details about the equation of motion, the numerical solver used, and the generation of noisy data. Finally, Sec.~\ref{ssec:cv} discusses how a cross-validation scheme is implemented in this analysis.

% Outline:
% \begin{itemize}
%     \item \textbf{Sparse Identification of Nonlinear Dynamics.} Discuss the SINDy method and where it came from, what it's based upon. Walk through the math of regression and discuss regularization and how it fits into least squares.
%     \begin{itemize}
%         \item \textbf{Weak SINDy method} Discuss how the integral approach of weak SINDy improves upon the strong form, and then walk through some of the math.
%         \item \textbf{Other SINDy methods} Discuss other methods of interest to physics, such as Lagrangian or Hamiltonian SINDy, trapping SINDy, and other modifications
%     \end{itemize}
%     \item \textbf{Simple simulation/proof-of-concept.} To verify the use of SINDy to discover interaction potentials, we use a simple initial value problem solver for the isotropic Yukawa interaction of two and three dust particles. Discuss how these were simulated and how noise was added.
%     \item \textbf{Cross-validation of data}. Discuss how cross-validation was employed to deduce the correct statistically learned model. Explain in-depth how either $k$-fold or leave-$p$-out cross-validation was used.
% \end{itemize}

\subsection{\label{ssec:sindy}The Sparse Identification of Nonlinear Dynamics (SINDy)}
The SINDy method of Ref.~\onlinecite{brunton_discovering_2016} is a data-driven analysis method that exploits the fact that most naturally occurring laws which govern dynamical systems are \textit{parsimonious}, or that they consist of relatively few terms.
The method uses sparse regression to find a differential equation that describes the underlying dynamics in the fewest number of terms possible. 
This sparse formulation has many advantages over previous methods that were used, such as the genetic symbolic regression \cite{schmidt_distilling_2009}, which is computationally expensive and does not scale well with system dimensionality.

SINDy is formulated to discover dynamical equations of the very general form
\begin{equation}
    \label{eqn:dyn}
    \tder{\mathbf{x}(t)} = \mathbf{f}(\mathbf{x}(t)), % used user-defined command for time derivative, see "User-defined functions" above
\end{equation}
where the vector $\mathbf{x}(t)=(x_1(t)\:x_2(t)\dots \: x_n(t))^\top$ is the state of the system at time $t$, $\{x_1 \dots x_n\}$ are the user-defined features used to describe the system, and $\mathbf{f}(\mathbf{x}(t))$ is a function that describes the system dynamics. 
Note that higher-order differential equations can be encoded in this ``first-order form'' by introducing more equations with dummy variables. For example, the second-order harmonic oscillator equation $\ddot{x}=\omega^2x$ can be represented by the first-order equations $\dot{x}=v,\ \dot{v}=\omega^2x$ so that $\mathbf{x}=(\:x\hspace{6pt}v\:)^\top$ and $\mathbf{f}=(\:v\hspace{8pt}\omega^2x\:)^\top$ from Eq.~\ref{eqn:dyn}.

Now, we further assume that the components of the function $\mathbf{f}$ from Eq.~\ref{eqn:dyn} can be written in the form
\begin{equation}
    \label{eqn:f}
    f_i(\mathbf{x}(t))=
    \sum_{j=1}^{p}c_{ij}\phi_j(\mathbf{x}(t)),
\end{equation}
where $\{\phi_j\}_{j\in[1,p]}$ is a set of $p$ nonlinear basis functions multiplied by constant coefficients $c_{ij}$. 
This will allow us to perform a \textit{linear} regression to determine the values of the $c_{ij}$ and express $\mathbf{f}$ as a linear combination of nonlinear functions $\phi_j$.

Since real data are sampled at discrete time intervals, we define the matrix $\mathbf{X}\in\mathbb{R}^{m \times n}$,
\begin{equation}
    \label{eqn:X_mat}
    \textbf{X} =
    \begin{pmatrix}
        \vert & \vert &        & \vert \\
        x_1   & x_2   & \cdots & x_n   \\
        \vert & \vert &        & \vert  
    \end{pmatrix},
\end{equation}
in lieu of the continuous function $\mathbf{x}(t)$. 
Here, the $i$th column is populated by the full history of $m$ discrete measurements of the system feature $x_i$, and there are $n$ number of such features. 
In a similar way, we define a ``library of terms'' $\boldsymbol{\Phi}\in\mathbb{R}^{m \times p}$,
\begin{equation}
    \label{eqn:Phi_mat}
    \boldsymbol{\Phi}(\mathbf{X}) =
    \begin{pmatrix}
        \vert              & \vert              &        & \vert \\
        \phi_1(\mathbf{X}) & \phi_2(\mathbf{X}) & \cdots & \phi_p(\mathbf{X})   \\
        \vert              & \vert              &        & \vert  
    \end{pmatrix},
\end{equation}
where the $i$th column is the function $\phi_i$ calculated at each time step using the data in $\mathbf{X}$. 
Finally, we define a matrix of coefficients $\mathbf{C}\in\mathbb{R}^{n \times p}$. 
Thus, we may now re-express Eq.~\ref{eqn:dyn} in matrix form as
\begin{equation}
    \label{eqn:dyn_mat}
    \mathbf{\dot{X}} = \boldsymbol{\Phi}(\mathbf{X})\mathbf{C}^\top.
\end{equation}

The SINDy regression seeks to minimize the loss $L$, which by construction balances minimizing error with minimizing the number of terms in the final equation. 
We define $L$ as
\begin{equation}
    \label{eqn:los}
    L=\norm{\mathbf{\dot{X}} - \boldsymbol{\Phi}(\mathbf{X})\mathbf{C}^\top}_2^2
    + \lambda R(\mathbf{C}),
\end{equation}
where the first term on the right hand side is the least squares loss according to Eq.~\ref{eqn:dyn_mat} and the second term is referred to as the ``sparsity prior.'' 
The function $R(\mathbf{C})$ imposes some constraint on the matrix $\mathbf{C}$ in order to promote sparsity in the final solution, and $\lambda$ is a hyperparameter which sets the strength of the sparsity promotion. 
In an ideal case $R(\mathbf{C})=\norm{\mathbf{C}}_0$, where $\norm{\cdot}_0$ is the $l_0$-norm and is simply a count of the nonzero terms in the matrix. 
As it happens, however, the $l_0$-norm is a non-convex function and therefore difficult to optimize. 
Therefore, $R(\mathbf{C})$ can be replaced with a number of other norms such as the $l_1$-norm (LASSO regression), the $l_2$-norm (Tikhonov regularization or ridge regression), or $L$ can be modified entirely to change the problem as in Sparse Relaxed Regularized Regression \cite{zheng_unified_2019}. 

In this work, we use the Sequential Thresholded Least Squares (STLSQ) approach \cite{brunton_discovering_2016}. 
We chose this approach for its simplicity---it performs a normal least squares regression ($R(\mathbf{C})=0$), sets to zero all least-squares coefficients which are below a predefined, fixed threshold, and performs the least squares again on the remaining coefficients. 
This process is repeated until no coefficients are below the threshold.
In terms of Eq.~\ref{eqn:los}, this amounts to using $\lambda=0$ and adding the additional step of thresholding outside of the loss function.
Note that the parameter $\lambda$ is set to zero for all applications using STLSQ, and that the STLSQ threshold and $\lambda$ are not the same.
Here, the threshold is the hyperparameter, which is set by the user. 
In Sec.~\ref{sec:res}, we vary the threshold hyperparameter to see what kinds of models SINDy identifies for different thresholds.

Here, we implement SINDy using the weak (or integral) formulation \cite{reinbold_robust_2021, messenger_weak_2021}. 
This approach leverages numerical integration and product rule to express Eq.~\ref{eqn:dyn} in weak form. 
This is done by multiplying both sides of the equation by a differentiable test function $\psi(t)$ such that $\psi\to0$ at the boundaries of the integration domain (or the total time interval $T$ during which measurements were taken). 
Thus, we can write
\begin{equation}
    \label{eqn:wea}
    -\int_0^T \frac{d\psi(u)}{du}\ \mathbf{x}(u)\ du =
    \int_0^T \psi(u)\ \mathbf{f}(\mathbf{x}(t))\ du,
\end{equation}
where $u$ is the integration variable. 
The loss function (Eq.~\ref{eqn:los}) can now be written as an integral and solved numerically without having to differentiate the data matrix $\mathbf{X}$. 
This approach drastically improves SINDy's robustness to noisy data, since numerically differentiating noisy data can be a large source of error.

In practice, this temporal integration is done by splitting the full time domain up into $K$ subdomains centered at random points $t_i$. 
The integration is then performed on the interval $[t_i - H_t, t_i + H_t]$, where $H_t$ is the half-width of the integration domain. More information on how this is done can be found in Ref.~\onlinecite{reinbold_robust_2021}, and the specific choice of parameters used here is discussed in Sec.~\ref{sec:res}.

Since the publication of Ref.~\onlinecite{brunton_discovering_2016}, SINDy has been improved upon in many different ways and adapted for many different applications. For example, it has been used to discover partial differential equations \cite{reinbold_robust_2021,rudy_data-driven_2017} in fluid dynamics and Lagrangians and Hamiltonians \cite{chu_discovering_2020,purnomo_sparse_2023} for simple mechanical systems like pendula.
Many of these applications are readily implemented within the \href{https://github.com/dynamicslab/pysindy}{PySINDy code} \cite{kaptanoglu_pysindy_2022}, which we utilize here for all analysis.

\subsection{\label{ssec:iso}Isotropic Yukawa simulation}
\begin{table}
    \caption{\label{tab:values}Values used to calculate the scaling constant $A$ (see Eq.~\ref{eqn:Fij_nodims}). Values come from Ref.~\onlinecite{pustylnik_plasmakristall-4_2016} or are approximate realistic values for a PK-4 dusty plasma with 3.38$\mathrm{\mu m}$ diameter particles.}
    \begin{ruledtabular}
        \begin{tabular}{llll}
        variable       & name                  & unit  & value \\
        \hline
        $k_B T_i$      & ion temperature       & eV              & $2.526\times10^{-2}$ \\
        $n_i$          & ion density           & $\text{m}^{-3}$ & $2.810\times10^{14}$ \\
        $n_d$          & dust density          & $\text{m}^{-3}$ & $1.000\times10^{11}$ \\
        $q_d$          & dust charge           & $e$             & $1.000\times10^4$ \\
        $m_d$          & dust mass             & kg              & $3.030\times10^{-14}$ \\
        $\lambda_{Di}$ & ion Debye length      & m               & $7.048\times10^{-5}$ \\
        $\omega_{pd}$  & dust plasma frequency & rad/s           & $9.781\times10^2$  \\
        % C\footnote{Some tables require footnotes.}
        %   &C\footnote{Some tables need more than one footnote.}
        %   & 12537.64 \\
        \end{tabular}
    \end{ruledtabular}
\end{table}

As mentioned in Sec.~\ref{ssec:ML}, the SINDy method has not been applied within the field of dusty plasma to the best knowledge of the authors. 
Therefore, we seek to validate the use of SINDy for a dusty plasma-relevant system using the isotropic Yukawa (shielded Coulomb) interaction potential. 
For this, we choose a system of 2 particles in two-dimensional (2D) space interacting with a Yukawa interaction potential 
\begin{equation}
    \label{eqn:yuk}
    \varphi(r)=\frac{1}{4\pi\varepsilon_0}\frac{q_d}{r}e^{-r/\lambda_{Di}},
\end{equation}
where $r$ is the distance from a dust particle (assumed to be a point particle) to a field point, and $q_d$ is the dust charge.
Note that this formulation with one spatial dimension is fully general for particles in up to three spatial dimensions.
This yields the following force between a pair of identical particles
\begin{equation}
    \label{eqn:Fij}
    \mathbf{F}_{ij} = 
    \frac{q_d^2}{4\pi\varepsilon_0}
    \left[
        \frac{1}{\lambda_{Di} r_{ij}}e^{-r_{ij}/\lambda_{Di}} +
        \frac{1}{r_{ij}^2}e^{-r_{ij}/\lambda_{Di}}
    \right] \hat{e}_{r_{ij}}
\end{equation}
where $r_{ij}$ is the magnitude of the vector $\mathbf{r}_{ij}=\boldsymbol{\rho}_i-\boldsymbol{\rho}_j$, $\boldsymbol{\rho}_k$ is the position of the $k$\textsuperscript{th} particle with respect to the origin, and $\hat{e}_{r_{ij}}=\mathbf{r}_{ij}/r_{ij}$ is the unit vector pointing from particle $j$ to particle $i$. 
We can normalize the quantities in this equation as follows, using a ``\textasciitilde'' to indicate normalized values. 
We normalize the interparticle spacing $\mathbf{r}_{ij}$ using the ion Debye length $\lambda_{Di}=(\varepsilon_0 k_B T_i/n_i e^2)^{1/2}$, yielding $\tilde{\mathbf{r}}_{ij}=\mathbf{r}_{ij}/\lambda_{Di}$ and time using the dust plasma frequency $\omega_{pd}=(n_d q_d^2/m_d \varepsilon_0)^{1/2}$, yielding $\tilde{t}=t\omega_{pd}$. 
To obtain the values of $\lambda_{Di}$ and $\omega_{pd}$, we use Plasmakristall-4 (PK-4)-relevant parameters, shown in Tab.~\ref{tab:values}. 
Using dimensionless quantities, the equation of motion of two dust particles (labeled i and j) interacting with the force $\mathbf{F}_{ij}$ becomes
\begin{equation}
    \label{eqn:Fij_nodims}
    \frac{d^2\tilde{r}_{ij}}{d\tilde{t}^2}
    =
    A
    \left[
        \frac{1}{\tilde{r}_{ij}} e^{-\tilde{r}_{ij}} +
        \frac{1}{\tilde{r}_{ij}^2} e^{-\tilde{r}_{ij}}\right]
\end{equation}
where $A$ is our ``scaling constant'' defined as $A=2q_d^2/(4\pi\varepsilon_0m_d\lambda_{Di}^3\omega_{pd}^2)$. 
Note that the equation of motion here is scalar-valued, a simplification which is only possible in the case of two bodies.

Synthetic data are generated by numerically solving Eq.~\ref{eqn:Fij_nodims} using the Livermore Solver for Ordinary Differential equations with Automatic stiffness adjustment \cite{petzold_automatic_1983} (LSODA), implemented through the \verb|integrate.solve_ivp| function in SciPy.
Two hundred trajectories are generated in this way with randomly (Gaussian) distributed initial velocities and separations. 
Gaussian-distributed noise is then added to the data with varying standard deviations $\sigma$ to simulate different amounts of error in real experimental data.
This approach comes from the PySINDy tutorial notebooks \cite{kaptanoglu_pysindy_2022} and is adapted here for a dusty plasma system.
In order to simulate and fit with PySINDy the second-order equation of motion seen in Eq.~\ref{eqn:Fij_nodims}, we must first convert it to first order form.
To do this, we introduce the simplified notation $r$ as a stand-in for $\tilde{r}_{ij}$ and add an additional equation with the variable $v\equiv\dot{r}$, where we use the overdot notation to represent time derivatives with respect to $\tilde{t}$.
We can now express Eq.~\ref{eqn:Fij_nodims} as
\begin{subequations}
    \label{eqn:1st_order_eom}
    \begin{align}
        \dot{r} &= v \label{eqn:1st_order_eom_rdot}\\
        \dot{v} &= A\left[
            \frac{e^{-r}}{r} +
            \frac{e^{-r}}{r^2}
        \right] \label{eqn:1st_order_eom_vdot}.
    \end{align}
\end{subequations}
The trajectories generated with this equation are then used as training data for SINDy to identify the equation of motion given in Eq.~\ref{eqn:Fij_nodims}.

To determine the SINDy library, we extrapolate based on the terms needed to form Eq.~\ref{eqn:1st_order_eom} and form a library $\boldsymbol{\Phi}$ as given by Eq.~\ref{eqn:Phi_mat}. In particular, the basis used is
\begin{equation}
    \label{eqn:lib}
    \{\phi_i\} = \left\{
        r, v, \frac{e^{-r}}{r}, \frac{e^{-v}}{v}, \frac{e^{-r}}{r^2}, \dots, \frac{e^{-v}}{v^4}
    \right\},
\end{equation}
where the unphysical $v$-dependent terms were included deliberately to test if SINDy would ignore them in favor of the $r$-dependent terms, and the linear terms were included to be able to fit Eq.~\ref{eqn:1st_order_eom_rdot}. This custom library was used in all analysis for this two-body system.

% Analysis is also extended to a system of three simulated dust particles interacting with a pure Yukawa potential. 
% This generates a system of three differential equations given by
% \begin{equation}
%     \label{eqn:3body}
%     \frac{d^2\tilde{\mathbf{r}}_{ij}}{d\tilde{t}^2}
%     =
%     A\left[2\mathbf{g}(\tilde{\mathbf{r}}_{ij}) - 
%      \mathbf{g}(\tilde{\mathbf{r}}_{jk}) - 
%      \mathbf{g}(\tilde{\mathbf{r}}_{ki})\right]
% \end{equation}
% where $i$, $j$, and $k$ represent the particle indices $\{0,1,2\}$ and are chosen such that $\epsilon_{ijk}=+1$, yielding three vector-valued differential equations that fully describe the system\footnote{$\epsilon_{ijk}$ here is the Levi-Civit\`a symbol, and for it to be equal to $+1$ in this case, we must have $ijk=\{012,120,201\}$.}. 
% The function $\mathbf{g}$ is vector-valued and multivariate; expressed as a function of the arbitrary vector $\mathbf{u}$, we have
% \begin{equation}
%     \label{eqn:fij}
%     \mathbf{g}(\mathbf{u})=
%     \mathbf{u}\left(
%         \frac{e^{-u}}{u^2}+\frac{e^{-u}}{u^3}\right)
% \end{equation}
% where we continue to follow the convention $u=|\mathbf{u}|$. 
% We simulate the three-body system in two dimensions by again solving the equations of motion, now a system of 6 coupled scalar-valued 2\textsuperscript{nd}-order differential equations. 
% After putting the equations in 1\textsuperscript{st}-order form as in Eq.~\ref{eqn:dyn}, we have 12 coupled scalar-valued differential equations.

\subsection{\label{ssec:cv}Cross-validation}
\begin{figure}
    \centering
    \includegraphics[width=\linewidth]{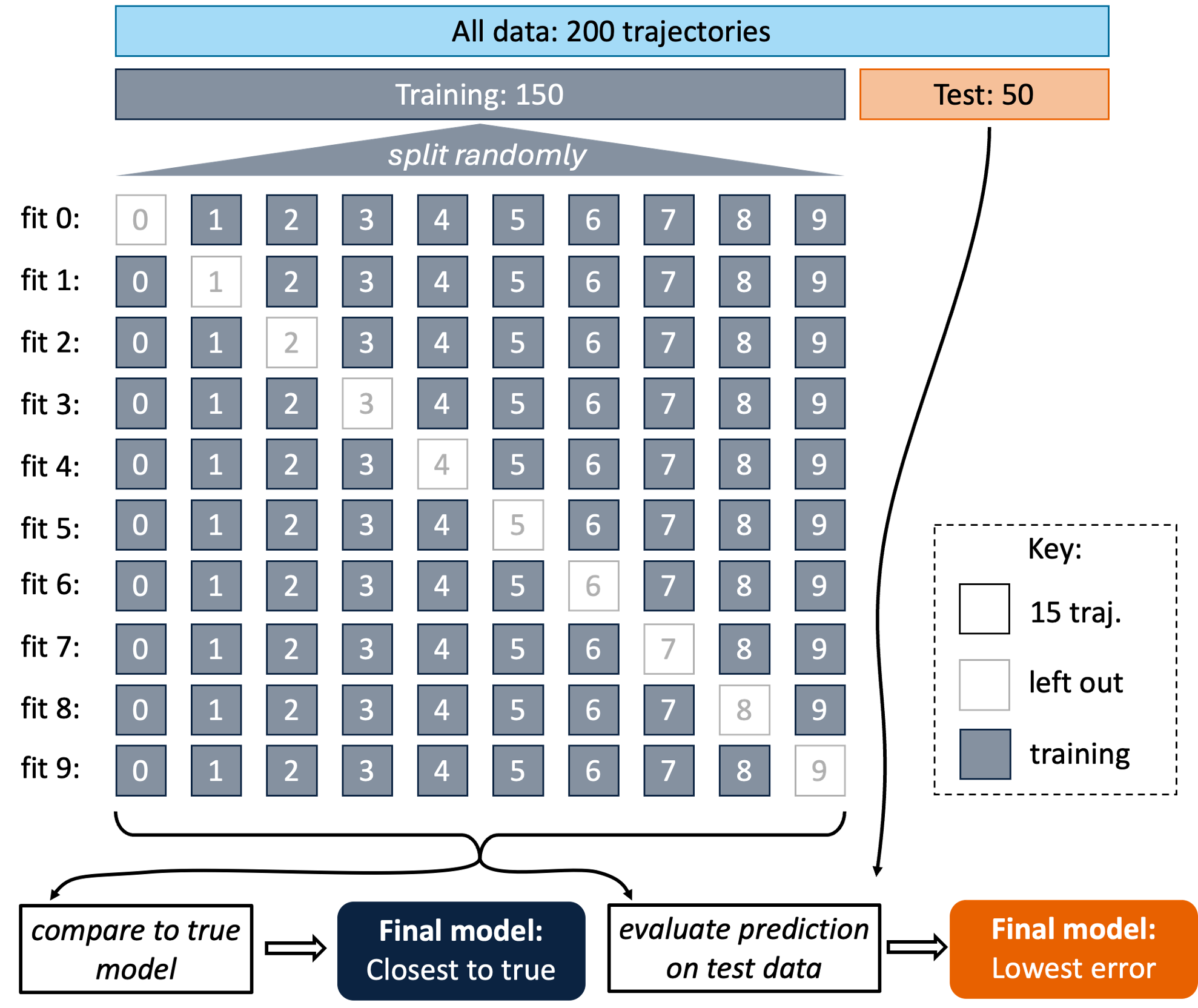}
    \caption{Visual representation of the $k$-fold cross-validation process (with $k=10$) used in the present analysis. The data (200 trajectories) are divided up into two groups: 150 (75\%) are used as training data and 50 (25\%) are used for testing. The training data are further divided into 10 ``folds,'' one of which is left out for each successive fit. Learned models can then be compared to the true model and evaluated on the test dataset.}
    \label{fig:cross-val}
\end{figure}

Cross-validation is an essential step in data-driven model generation, as it minimizes the possibility of outliers or abnormal parts of the data causing inaccuracies in the learned model.
In the present study, we implement the 10-fold cross-validation scheme, which is depicted in Fig.~\ref{fig:cross-val} and can be described as follows. 
First, the 200 trajectories are broken up into two parts, 50 (25\%) for \textit{testing} and 150 (75\%) for \textit{training}. 
Then, the training data is broken up into 10 groups of 15 trajectories each, referred to as ``folds.'' 
A SINDy model is trained using all but one fold, which is left out of training. 
This is repeated so that all of the training data has been left out of training at some point.

\begin{figure*}
    \centering
    \includegraphics[width=0.8\linewidth]{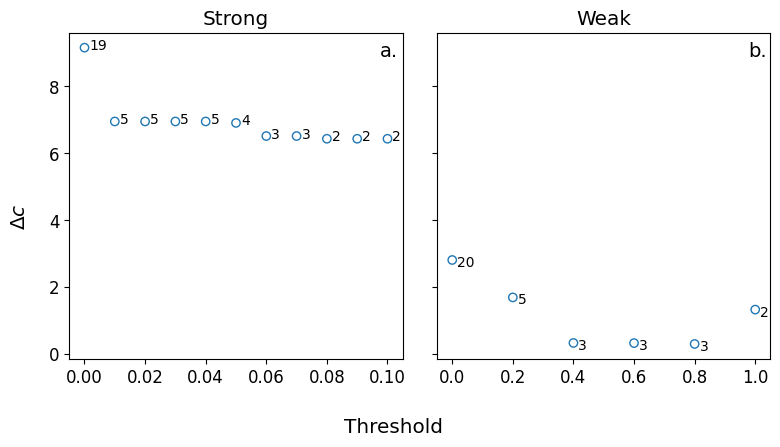}
    \caption{Coefficient deviation of successive learned SINDy models (Eq.~\ref{eqn:coef_dev}) plotted as a function of the STLSQ threshold parameter used during fitting. Data used here has a noise level of $0.1 \lambda_{Di}$ and each point is labeled with the number of terms in the model; note that the true model (Eq.~\ref{eqn:1st_order_eom}) has three terms.}
    \label{fig:wkvsstr}
\end{figure*}

Each of the 10 models generated during cross-validation are evaluated in two ways. 
First, the learned model coefficient matrix $\mathbf{C}_\mathrm{learned}$ is compared to the coefficient matrix of the true model $\mathbf{C}_\mathrm{true}$, given by Eq.~\ref{eqn:Fij_nodims}. 
This is done by calculating the ``coefficient deviation'' $\Delta c$, which is written as
\begin{equation}
    \label{eqn:coef_dev}
    \Delta c = \sum_{i,j}|c_{ij\ \mathrm{learned}}-c_{ij\ \mathrm{true}}|,
\end{equation}
where the constants $c_{ij}$ are the elements of the matrices $\mathbf{C}$. In the case that the true model is not known, however, Eq.~\ref{eqn:coef_dev} cannot be used. 
Thus, we also compare the learned model's prediction of the derivative $\dot{\mathbf{X}}_\mathrm{pred}$ to the calculated finite-difference derivative using the test dataset $\dot{\mathbf{X}}_\mathrm{calc}$. 
This is done by simply calculating the root mean squared error $\epsilon$,
\begin{equation}
    \label{eqn:score}
    \epsilon = ||\dot{\mathbf{X}}_\mathrm{pred}-\dot{\mathbf{X}}_\mathrm{calc}||_2.
\end{equation}
\section{\label{sec:res}Results}

%%%%%%%%%%%%%%%%%%%%%%%%%%%%%%%%%%%%%%%%%%%%%%%%%%%%%
% RESULTS
%%%%%%%%%%%%%%%%%%%%%%%%%%%%%%%%%%%%%%%%%%%%%%%%%%%%%

\begin{figure}
    \centering
    \includegraphics[width=\linewidth]{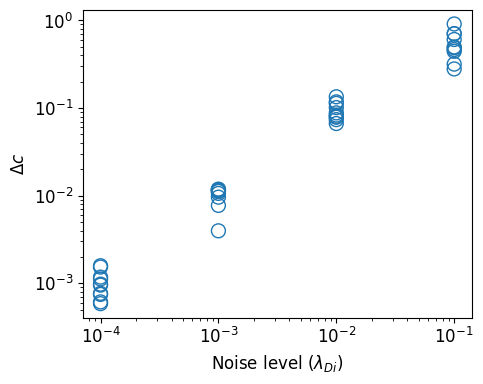}
    \caption{Coefficient deviations $\Delta c$ for models determined from the synthetic data using the weak SINDy formulation described in Sec.~\ref{sec:met}. Ten models were generated at each of the four noise levels tested here, each with a different random sampling of temporal data.}
    \label{fig:robustness}
\end{figure}

In this section, we evaluate the ability of this approach to recover the known model given in Eq.~\ref{eqn:Fij_nodims}. 
To do this, we sweep through several threshold values and use the coefficient deviation metric $\Delta c$ defined in Eq.~\ref{eqn:coef_dev} to evaluate the effectiveness of SINDy to determine the true coefficients. 
First, the ``strong'' (no integral) SINDy approach is compared to the weak (integral) formulation at two noise levels (see Sec.~\ref{ssec:sindy}). The weak formulation is found to perform better with higher noise magnitude than the strong approach, as shown in Fig.~\ref{fig:wkvsstr}. 
Next, the noise robustness of the weak approach is determined by testing the recovery of coefficients at higher noise levels, as shown in Fig.~\ref{fig:robustness}.
To determine the effectiveness of the SINDy approach, the cross-validation routine described in Sec.~\ref{ssec:cv} is performed for a range of different STLSQ threshold parameter values. 
For the remainder of the results presented here, this threshold parameter scan is used to choose an optimal threshold for a given level of noise.

For low noise magnitudes (e.g. $\sigma=10^{-4}\lambda_{Di}$), strong-form SINDy is favorable to use since it is faster \footnote{With these parameters and amount of data, the strong approach generally takes less than a minute while the weak approach can take up to two minutes.} and essentially performs as good as the weak approach. 
The two methods start to differ at higher levels of noise, for example, the $\sigma=0.1\lambda_{Di}$ case being shown in Fig.~\ref{fig:wkvsstr}. 
Here, the coefficient deviation $\Delta c$ of successive models is plotted against the STLSQ threshold used to generate the model and each point is labeled with the number of terms in the learned equations. 
Note that the true model, Eq.~\ref{eqn:1st_order_eom}, has 3 terms after converting it to first-order form. 
As evidenced by the high value of coefficient deviation in Fig.~\ref{fig:wkvsstr}a, the models identified by the strong approach that have the same number of terms as the true model differ qualitatively and quantitatively from it. For example, the model discovered with a threshold of 0.06 is
\begin{subequations}
    \label{eqn:strong_0_06}
    \begin{align}
        \dot{r} &= 0.994\ v \\
        \dot{v} &= 0.017\ A\ v + 0.586\ A\ \frac{e^{-r}}{r} \label{seqn:strong_vdot},
    \end{align}
\end{subequations}
where $r$ is simplified notation for the $\tilde{r}_{ij}$ introduced in Eq.~\ref{eqn:Fij_nodims}, the dot represents a derivative with respect to $\tilde{t}$, $A$ is the scaling constant of Eq.~\ref{eqn:Fij_nodims}, and the variable $v$ is introduced to transform Eq.~\ref{eqn:Fij_nodims} into first-order form shown in Eq.~\ref{eqn:1st_order_eom}.
As we can see here, the model discovered includes a nonphysical velocity-dependent term in Eq.~\ref{seqn:strong_vdot} and the value of the coefficient on the correct rational term is off by \textasciitilde50\%.
At the higher threshold of 0.08, the nonphysical velocity-dependent term is removed by the thresholding, yielding a model qualitatively similar to the true model but quantitatively quite different. 
This can be compared to the weak model at a threshold of 0.4,
\begin{subequations}
    \label{eqn:weak_0_4}
    \begin{align}
        \dot{r} &= 0.999\ v \\
        \dot{v} &= 0.967\ A\ \frac{e^{-r}}{r} + 0.962\ A\ \frac{e^{-r}}{r^2},
    \end{align}
\end{subequations}
which recovers the true model to much higher accuracy.
This result is in agreement with previous work utilizing weak SINDy \cite{reinbold_robust_2021,kaptanoglu_pysindy_2022,messenger_weak_2021}.

For the weak method, two key parameters are at play: the number of temporal subdomains $K$ and the subdomain half-width $H_t$, as mentioned in Sec.~\ref{sec:met}. 
In general, the weak formulation performs better with higher $K$-values, as it adds points to the regression \cite{kaptanoglu_pysindy_2022}. 
For our application, we used $K=500$, as that increased the noise robustness a significant amount while not increasing the computational time too much. 
We suspect that an even higher value would improve the performance, albeit marginally. For the subdomain half-width, we used the PySINDy default value of $T/20$, where $T$ is the length of the total time domain.

The noise robustness of the approach was also investigated by performing the weak-form SINDy analysis with datasets at four different noise levels, the results of which are shown in Fig.~\ref{fig:robustness}. 
These results were generated by first doing a threshold scan as shown in Fig.~\ref{fig:wkvsstr} to determine an optimal threshold to use at each noise level being tested.
There is some degree of subjectivity in this process as the threshold is chosen based on plots such as Fig.~\ref{fig:wkvsstr}.
Once a threshold had been determined, cross-validation was performed on the same dataset of 200 trajectories with different levels of noise added 10 different times.
The primary source of the differences in the learned models is the different random sampling of subdomains that weak SINDy uses each time the analysis is performed.
For higher noise level, the accuracy of the learned coefficients is decreased, but the models are in qualitative agreement with the true model. 
In general, we found that the true model can be recovered for noise up to 0.2 $\lambda_{Di}$. 
One problem with many of the learned models using this method is that some extraneous terms with small coefficients are not effectively eliminated by the STLSQ optimizer and remain in the final solution. 
For example, in a case with $\sigma = 0.1\lambda_{Di}$, the following model is learned with a threshold of 0.4:
\begin{subequations}
    \begin{align}
        \dot{r} &= 0.998\ v \label{seqn:weak_rdot}\\
        \dot{v} &= 0.949\ A\ \frac{e^{-r}}{r} + 0.943\ A\ \frac{e^{-r}}{r^2} + 0.07319\ A\ \frac{e^{-r}}{r^3}\label{seqn:weak_vdot}.
    \end{align}
\end{subequations}
This is inconsistent with the threshold used, since it is larger than the smallest coefficient in Eq.~\ref{seqn:weak_vdot}. This is either due to the STLSQ optimizer reaching the max number of iterations or related to how the weak-form library interfaces with the optimizer. When this is the case, increasing the threshold eliminates the one remaining term in Eq.~\ref{seqn:weak_rdot} before the small-magnitude extraneous term in Eq.~\ref{seqn:weak_vdot}.

%%%%%%%%%%%%%%%%%%%%%%%%%%%%%%%%%%%%%%%%%%%%%%%%%%%%%
% DISCUSSION
%%%%%%%%%%%%%%%%%%%%%%%%%%%%%%%%%%%%%%%%%%%%%%%%%%%%%
\section{\label{sec:dis}Discussion}
% Partial(?) outline of discussion section
% \begin{itemize}
%     \item Interpretation of results
%     \item Limitations
%     \item Extensions and future work
%     \begin{itemize}
%         \item Many-body systems
%         \item Anisotropic potentials
%         \item Nonreciprocal systems
%     \end{itemize}
% \end{itemize}

\subsection{\label{ssec:interp}Interpretation of results}
These results show the determination of the equation of motion of a two-body Yukawa system using data from a initial value problem solver with added Gaussian-distributed noise. Two formulations were compared, the strong form and the weak form of SINDy.
The weak SINDy formulation was able to reproducibly determine the form of the equation with accuracy of $\Delta c<1$ for noise levels of up to 0.1 $\lambda_{Di}$, or $\approx7\mu m$ in the PK-4 experiment.
Returning to the PK-4 experiment for reference, the value of $\lambda_{Di}\approx70\mu m$ reported in Tab.~\ref{tab:values} corresponds to approximately 5 pixels in the particle observation cameras of that experiment (pixel size $\approx14\mu m$, typical dust diameter $d\approx3.8\mu m$), which means that the noise level of $0.1\lambda_{Di}$ corresponds to half a pixel. Commonplace particle tracking methods which are frequently employed within the field of dusty plasma \cite{crocker_methods_1996,sbalzarini_feature_2005} are able to resolve particle motion to sub-pixel resolution given optimal parameters. 
This means that for relevant experimental conditions, the weak SINDy method could be used to probe experimental particle interaction potentials, opening the door to data-driven discovery of elusive mathematical forms for particles suspended in nonequilibrium plasmas.

Results using the weak SINDy approach vary largely because of the random selection of a subset of time domains from the full time domain.
Certain selections of subdomains perform better than others, which is in part why the cross-validation scheme is used, so the best of an ensemble of models can be chosen.
Results can be improved by using more subdomains sampled near the earlier time in the simulation, as this is when the particles have the greatest acceleration due to strongest interaction.
Once particles are farther away from each other, the weaker interaction provides less information that the regression can exploit to deduce which terms are more important, since the acceleration approaches zero as the distance between the particles approaches infinity.
Random points are more favorable for chaotic systems, however, which can only occur in systems of first-order differential equations if there are three or more equations.
In that case, the accelerations and velocities can vary wildly throughout the time domain and, thus, provide more information that would help deduce which terms from the library are the most important in describing the dynamics.
For dusty plasma, this condition would be met by the characteristics of many common experiments, i.e. with many-body systems like a cloud of dust particles suspended in plasma.
In fact, the simple inclusion of a third particle causes the new system of equations to exhibit chaos.
See Sec.~\ref{ssec:ext} for a more complete discussion on how this method can be applied to such a system.

\subsection{\label{ssec:sel}Model selection}
\begin{figure}
    \centering
    \includegraphics[width=\linewidth]{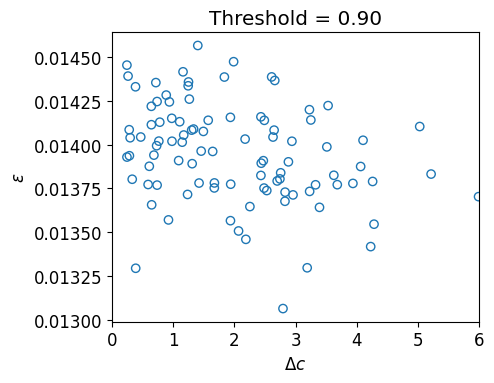}
    \caption{The prediction error $\epsilon$ plotted against the coefficient deviation $\Delta c$ of one hundred models generated during cross-validation. All the models shown here use an STLSQ threshold of 0.9, which corresponds to 0.19 $A$. There is no visible correlation between the two metrics $\Delta c$ and $\epsilon$, a linear regression analysis returning an $r^2$ coefficient of $10^{-10}$.}
    \label{fig:mdl_sel}
\end{figure}

The metric $\Delta c$ of Eq.~\ref{eqn:coef_dev} relies on knowledge of the true model and was used to determine the effectiveness of the weak SINDy approach with cross-validation in Sec.~\ref{sec:res}. 
However, in the case of an unknown model, when an attempt is being made to uncover a physical model directly from data, this metric cannot be used to select the best model during cross-validation. 
Instead, some other criterion must be used to select the model, the most obvious one being the error in Eq.~\ref{eqn:score}. 
In Fig.~\ref{fig:mdl_sel}, we plot the correlation between the coefficient deviation used in Sec.~\ref{sec:res} and the aforementioned error for one hundred weak models generated from performing cross-validation ten times.
The models plotted here all have a threshold of 0.9 $\lambda_{Di}$, which is on the higher end of thresholds that will generate the true model for this noise level, ensuring that outlier models with high $\Delta c$-values occur seldom.
From the figure, we see that there is no visible correlation between the model evaluation metrics, and linear regression analysis shows a Pearson correlation coefficients ($r^2$-value) on the order of $10^{-10}$. 

This means that the prediction accuracy as it is defined by Eq.~\ref{eqn:score} is \textit{not} a good metric to use for model selection. 
This metric was selected because of its use in the PySINDy tutorial notebooks \cite{kaptanoglu_pysindy_2022} to evaluate models, but for model selection in an unknown system this result does not give confidence that the actual ``true'' model could be determined from cross-validation. Future work should investigate other metrics, perhaps balancing sparsity with prediction power to provide a more reliable way to choose a model from the ensemble generated by cross-validation.

\subsection{\label{ssec:lim}Limitations}
The present study has three major limitations. First, it is important to note that we examine the two-body case, which is trivial compared to the much more interesting many-body case in 2D or 3D that arises in most dusty plasma experiments.
Another limitation of this study is that the dust interaction potential studied here is isotropic, whereas in many interesting research contexts we seek to learn an anisotropic potential. 
For instance, as mentioned in Sec.~\ref{sec:intro}, the PlasmaKristall-4 (PK-4) experiment has shown the presence of string-like dust structures that imply a high degree of anisotropy in the effective dust interaction potential.
Finally, an important aspect of certain dusty plasma environments is the nonreciprocity of the effective dust interaction potential, which is not addressed here. 
This effect was addressed very nicely in Ref.~\onlinecite{yu_physics-tailored_2025} using an artificial neural network, albeit without the interpretability and speed offered by the present approach.
Despite these limitations, through this proof-of-concept we aim to show that the SINDy method can be used to discover equations of motion directly from particle data, allowing for data-driven mathematical modeling of dusty plasma and paving the way for the analysis of many-body, anisotropic, and nonreciprocal cases using sparse regression to follow.

Another consideration is that the terms in the model library are highly correlated with one another, 
which is clear from the obvious qualitative similarities between rational expressions of different order. This makes it difficult for the regression to choose between library terms, especially for a very high level of noise.
In order to improve the performance of the SINDy method, it is favorable to use orthogonal (linearly independent) functions, or at least functions with large qualitative differences.
For instance, to apply this analysis to the anisotropic case mentioned above, the Legendre Polynomials could be used to describe the angular dependence of the potential structure.
These functions are of course orthonormal and have been used to describe the functional form of the dust interaction potential in Electrorheological plasmas \cite{kompaneets_potential_2007,ivlev_electrorheological_2010}.

\subsection{\label{ssec:ext}Extensions and future work}
The next immediate application of this work is to investigate systems of two bodies in which the background plasma environment is more accurately simulated. This can be done with the Dynamics Response of Ions And Dust (DRIAD) code \cite{matthews_dust_2020,vermillion_influence_2022}.
In this case, an angular dependence of the potential can be learned, allowing for the discovery of the interaction potential in an anisotropic plasma environment.
While there is a great deal of analytical derivation \cite{kompaneets_potential_2007,ivlev_electrorheological_2010}, computational simulation \cite{vermillion_influence_2022, matthews_effect_2021, vermillion_interacting_2024, mendoza_ion_2025}, and experimental analysis \cite{pustylnik_three-dimensional_2020, mitic_long-term_2021} investigating this anisotropy, our approach would be the first approach to learn this kind of potential directly from data.
This could offer unique insight into this system and aid in the discovery of new mathematical ways of describing the dynamics of dust grains in plasma.

Another extension of this work is applying it to experimental data obtained from dusty plasma experiments with few particles.\cite{liang_determining_2023, van_huijstee_probing_2025}.
In these experiments, motion of one or a few dust particles are tracked and used to recover information about the background plasma such as the nature of the sheath electric field.
With our method, it may be possible to recover an equation of motion for a dust particle in one of these environments and thus to better understand the physical mechanisms at play.

As mentioned in Sec.~\ref{ssec:lim}, many-body systems are more commonly encountered within the field of dusty plasma than few-body systems.
Indeed, the physics of these systems is harder to describe and, thus, less well-understood.
One possible approach to these systems is to use the mathematical machinery of mean field theory along with weak SINDy \cite{messenger_learning_2022}.
This direction looks at trajectories of many particles and infers an average equation of motion with separate terms for the pair potential, environmental potential, and stochastic effects, like Brownian motion. 
This approach could be coupled with the approach of Lagrangian SINDy \cite{chu_discovering_2020,purnomo_sparse_2023} to learn an average Hamiltonian or Lagrangian for the dust cloud.

Addressing the issue of non-reciprocity might also be possible with an extension of our approach. Owing to the work of Ivlev \textit{et al.} \cite{ivlev_statistical_2015}, an analytical basis describing the nonreciprocity in dusty plasma systems has been set.
Using weak SINDy to learn the pseudo-Hamiltonian \cite{ivlev_statistical_2015} for nonreciprocal systems is perhaps a first direction to take here.

%%%%%%%%%%%%%%%%%%%%%%%%%%%%%%%%%%%%%%%%%%%%%%%%%%%%%
% CONCLUSION
%%%%%%%%%%%%%%%%%%%%%%%%%%%%%%%%%%%%%%%%%%%%%%%%%%%%%
\section{\label{sec:con}Conclusion}
The use of the Sparse Identification of Nonlinear Dynamics \cite{brunton_discovering_2016} using the weak form has been used to identify an equation of motion for two particles from synthetic data has been presented. Leveraging the use of cross-validation, the method is able to learn equations of motion which approximate the true model Eq.~\ref{eqn:Fij_nodims} from a library of terms Eq.~\ref{eqn:lib} accurately but with the inclusion of extraneous terms which have small coefficients.

With some improvements, this approach could be used to learn interaction potentials of particles in more realistic simulations and experiments, possibly aiding in the identification of anisotropic potentials, non-reciprocal forces, or mean-field equations. As a non-neural-network machine learning method, using SINDy to discover equations from data represents a novel perspective on the scientific investigation of dusty plasma and could offer new interpretable insights into the physical processes at play in the system.

%%%%%%%%%%%%%%%%%%%%%%%%%%%%%%%%%%%%%%%%%%%%%%%%%%%%%
% ACKNOWLEDGEMENTS
%%%%%%%%%%%%%%%%%%%%%%%%%%%%%%%%%%%%%%%%%%%%%%%%%%%%%
\section{Acknowledgments}
Support for this work from NSF Grants 2308742, 2308743, EPSCoR FTPP OIA-2148653, and the US Department of Energy, Office of Science, Office of Fusion Energy Sciences under award number DE-SC0024547, DE-SC-0021334 is gratefully acknowledged.

%\nocite{*}
%\bibliographystyle{plain}
\bibliography{my_lib}
% Produces the bibliography via BibTeX.

\end{document}